\documentclass[10pt,preprintnumbers,pra,aps,
nofootinbib,tightenlines,floatfix, twocolumn,superscriptaddress,longbibliography]{revtex4-1}

\usepackage{hyperref}
\usepackage[all]{hypcap}
\usepackage{url}
\usepackage{amsmath,amssymb}
\usepackage{graphicx}
\usepackage{bm}
\usepackage[utf8]{inputenc}
\usepackage{csquotes}

\allowdisplaybreaks
\newcommand{\Tr}{\ensuremath{\text{Tr}}}

\newcount\hour \newcount\hourminute \newcount\minute
\hour=\time \divide \hour by 60
\hourminute=\hour \multiply \hourminute by 60
\minute=\time \advance \minute by -\hourminute
\newcommand{\mydate}{\ \today \ - \number\hour :\number\minute}

\begin{document}

\title{Hierarchical Qubit Maps and Hierarchical Quantum Error Correction}
\author{Natalie Klco}
\email{natklco@caltech.edu}
\affiliation{Institute for Quantum Information and Matter and Walter Burke Institute for Theoretical Physics,
California Institute of Technology, Pasadena CA 91125, USA}
\author{Martin J.~Savage}
\email{mjs5@uw.edu}
\affiliation{InQubator for Quantum Simulation (IQuS), Department of Physics, University of Washington, Seattle, WA 98195.}

\date{\mydate}
\preprint{IQuS@UW-21-010}


\begin{abstract}
We consider hierarchically implemented quantum error
correction (HI-QEC), in which the fidelities of logical qubits are differentially
optimized to enhance the capabilities of quantum devices in scientific applications.
By employing qubit representations that propagate hierarchies in simulated systems to those in
logical qubit noise sensitivities, heterogeneity in the distribution of physical-to-logical qubits can be systematically structured.
For concreteness, we estimate HI-QEC's impact on {\tt surface code} resources in computing low-energy observables to fixed precision,
finding up to $\sim 60\%$ reductions in qubit requirements plausible in early error corrected simulations.
Hierarchical qubit maps are also possible without error correction in qubit and qudit systems where fidelities are non-uniform, either unintentionally or by design.  Hierarchical optimizations are another element in the co-design process of quantum simulations for nuclear and particle physics.
\end{abstract}

\date{\mydate}
\maketitle

\section{Introduction}

Rapid advances in quantum technologies are stimulating significant activity toward
simulating quantum field theories (QFTs) and quantum many-body systems
with future quantum computers.
These advances are not only important in fundamental and applied research, but also in the development of quantum computers themselves.
Such simulations are expected to enable calculations of observables that cannot be accessed with sufficient precision at scale with classical computers, including  real-time dynamics of systems out of equilibrium and systems with large numbers of particles.

Currently in the NISQ-era~\cite{Preskill2018quantumcomputingin}, digital quantum devices available for scientific applications have
modest numbers of physical qubits, limited but improving fidelity,  noisy gate operations, and short coherence times.
Device noise in current simulations is mitigated to some degree by:
selecting a configuration with the highest fidelity qubits and desired entangling gates within a given quantum processing unit;
extrapolating entangling gate errors with global CNOT
replacement~\cite{PhysRevLett.119.180509,PhysRevX.8.031027,Dumitrescu:2018njn,Klco:2018kyo,Yeter-Aydeniz:2018mix,Mishra:2019xbh,Klco:2019xro,Gustafson:2019vsd,Roggero:2020sgd,Yeter-Aydeniz:2020jte,Ciavarella:2021nmj,Atas:2021ext,Yeter-Aydeniz:2021olz,Hall:2021rbv} and local stochastic insertions~\cite{Klco:2019evd,He:2020udd};
post-selection of physical subspaces~\cite{Martinez:2016yna,Klco:2019evd};
addressing measurement errors through inversion of simple noise models~\cite{ibmmeasurementerror}, classical conditional probabilities~\cite{Funcke:2020olv} and majority voting~\cite{Ciavarella:2021nmj};
and time-dependent in-vivo calibration workflows and \enquote{in-medium} gate correction~\cite{Klco:2019xro}.
Excitingly, first steps toward experimental demonstration of quantum error correction (QEC) have emerged, e.g., recent demonstrations of exponential convergence of the {\tt repetition code} ~\cite{2015Natur.519...66K,wootton2018repetition,Chen2021,egan2021faulttolerant}, error detection in the {\tt surface code}~\cite{Kitaev_2003,Gottesman:1997zz,freedman2001projective,Bravyi:1998sy,Dennis_2002,Wang:2002ph} on the 53-qubit {\tt Sycamore} superconducting processor~\cite{Google48651,Chen2021}, fault-tolerant operations in the 9-qubit {\tt Bacon-Shor code} on 13 trapped $^{171}Yb^+$ ions~\cite{egan2021faulttolerant}, and real-time error correction in the {\tt [[7,1,3]] color code} on ten trapped $^{171}Yb^+$ ions~\cite{ryananderson2021realization}.
The operational protocols of QEC introduce a level of abstraction through the transition from physical to logical qubits.
In the use of general purpose quantum computers designed without
preference to the local fidelity requirements of specific calculations, logical qubits are comprised of a uniform number of physical qubits, each subject to the same error-correction layouts and codes.
With this approach, there is a generic multiplicative scaling factor, increasing logarithmically with the desired precision, that converts from logical to physical qubit requirements.

Similar to the importance of designing quantum hardware with commensurate connectivity (and thus dimensionality) to the simulated system~\cite{DBLP:journals/qic/JordanLP14,doi:10.1137/18M1231511}, as profitably demonstrated in classical computing for, e.g., lattice
quantum chromodynamics (QCD)~\cite{Christ2011,Marinari1986},
it is equally important to design mappings to computational Hilbert spaces that recover the simulated Hilbert space structure, reflecting symmetries and scales.
As discussed below, there is utility in organizing a Hilbert space such that a hierarchy can be identified in the relative contributions of each logical qubit.
In simulating lattice QFTs using a Hilbert-space layout that mirrors the spatial distribution of fields, a further choice of digitization of the field at each spatially local site will determine whether a hierarchical identification of each logical qubit is possible.
For example in digitizing the scalar field~\cite{Jordan1130,DBLP:journals/qic/JordanLP14,Marshall:2015mna,somma2016quantum,PhysRevA.98.042312,Yeter-Aydeniz:2018mix,Klco:2018zqz,Barata:2020jtq,Macridin:2021uwn},
the Jordan-Lee-Preskill (JLP)~\cite{Jordan1130,DBLP:journals/qic/JordanLP14} basis of field eigenstates mapped in binary to each local Hilbert space differentially distributes \enquote{responsibility} of bands of frequency modes among the logical qubits.
An analogous hierarchy is also present when digitizing with a binary mapping of harmonic oscillator modes~\cite{Klco:2018zqz}, where logical qubits are \enquote{responsible} for generating amplitudes of harmonic oscillator states separated by exponentially differing magnitudes of the principle quantum number.
While hierarchical identification of logical qubits is expected to be broadly manifest in field digitizations, it is not guaranteed for all representations, e.g., lattice delocalized field operators.
An example of this diversity can be found in formulations for simulating Yang-Mills gauge theory (for further discussions, see Refs.~\cite{Banuls:2019bmf,Klco:2021lap}).
The benefits of hierarchically ordered qubits are expected to impact quantum simulations of scalar field theory, lattice gauge theories (including QCD), effective field theories~(EFTs), chemistry, and other domain sciences.
When a mapping to logical qubits is organized to manifest a hierarchy of energy scales, for example, both low-energy wavefunctions and low-energy observables will be dominantly sensitive to the fidelities of IR qubits, and differentially less sensitive to errors in the UV logical qubits.

By preferentially protecting the fidelities of IR logical qubits, quantum simulations depart from the computationally challenging goal of
computing all observables to the same fixed precision and toward the goal of
computing some observables with enhanced precision in an isolated energy regime of interest.
As one of many deviations from general purpose quantum computers envisaged for co-designing quantum devices for specific scientific applications, this preferential protection can be achieved through strategic distributions of physical qubits heterogeneously,  optimizing resources for the computation of select observables.
Therefore, for the same reasons that EFTs are useful in describing low-energy physics, the physical qubit requirements can be tuned to be less than simply a multiplicative factor times the logical qubit requirements.
Resulting  \enquote{surplus}  qubits  can be put to good use in increasing the dimensionality of local Hilbert spaces
 describing fields or variables
that have been truncated by finite resources.
This is particularly relevant in near- and intermediate-term quantum simulations.
For example, truncations of the local \enquote{color} Hilbert spaces in Yang-Mills gauge theories limit the approach to the spatial continuum,
where the electric-basis wavefunctions become increasingly delocalized~\cite{PhysRevD.11.395,PhysRevA.73.022328,Zohar:2012xf,Klco:2019evd,Banuls_2020,Davoudi:2020yln,Haase:2020kaj,Ciavarella:2021nmj}.
An additional qubit each in the p- and q-space registers describing an SU(3) Yang-Mills gauge field increases the
number of color states in the digitized gauge-field link variables included in the simulation by a factor of 32 asymptotically.

This
work illustrates and estimates the impact of heterogeneity in the physical-to-logical qubit distributions on the calculation of low-energy observables of smooth wavefunctions.
While the asymptotic savings are modest, those relevant to the transition period from NISQ to fault tolerance, where physical error rates are below but near error correction thresholds, are expected to expand the scientific capabilities of available devices.
We provide detailed examples to demonstrate this optimization subject to depolarizing noise on a Gaussian wavefunction, examining frequency hierarchies through Walsh-Hadamard transforms (WHT).
With the simulated performance of {\tt surface code} error correction, we find that physical qubit requirements can be reduced by up to $\sim 60\%$ through hierarchical redistribution.

\begin{figure}
  \includegraphics[width=0.95\columnwidth]{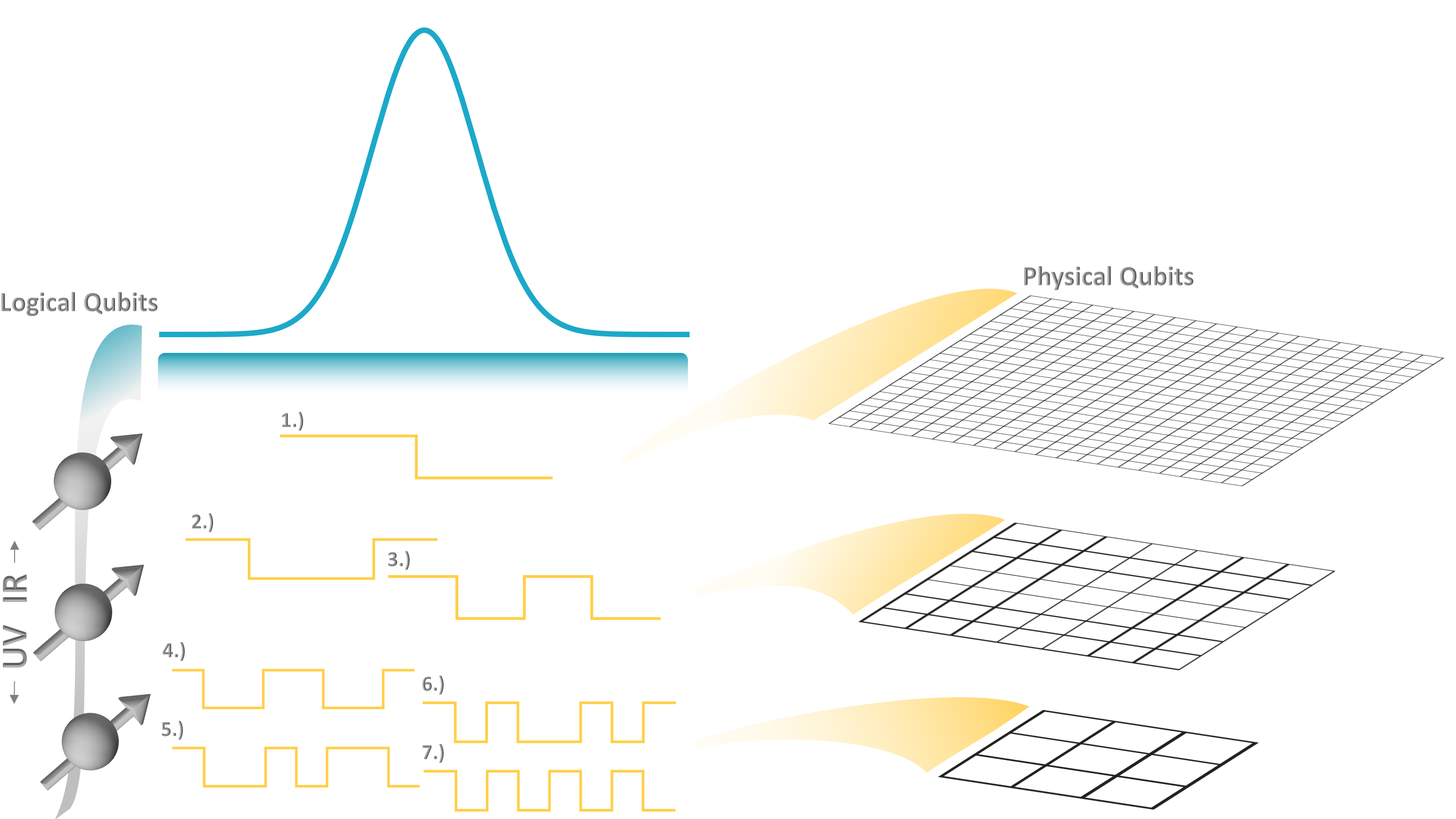}
  \caption{Diagrammatic expression of HI-QEC for logical qubits embedded in e.g., {\tt surface codes}.
  Organizing wavefunction digitization such that sequency bands, the digital analog of frequency, are generated
  by rotations of individual qubits (independently of more-UV qubits)
  allows for heterogeneous optimization of the physical qubit distribution,
  protecting the logical qubits preferentially at scales of interest,
  e.g., the IR as depicted.
  }
  \label{fig:diagram}
\end{figure}

\section{Basis for Observables}
Consider observables measured in their eigenbasis,  represented by real diagonal matrices.
Such observables may be decomposed into an orthonormal basis of Pauli operators, $\mathcal{O}_j $,  as,
\begin{equation}
  \mathcal{O} = \sum_{j=0}^{2^n-1} \beta_j  \mathcal{O}_j \qquad , \qquad \mathcal{O}_j = \bigotimes_{b=0}^{n-1}  \sigma_{(3j_b)}
  \ \ \ ,
  \label{eq:opdef}
\end{equation}
where $\sigma_0 = \mathbb{I}$, $\sigma_3 = Z$,
and the binary representation of $j = \{j_{n-1},  \cdots j_0 \}$ with $j_b \in \{0, 1\}$
has been used to define a tensor product operator of distributed Pauli-$Z$s in the location of the binary ``1''s.~\footnote{
For example, in a four-qubit system, $\mathcal{O}_{9} = Z\otimes \mathbb{I} \otimes \mathbb{I} \otimes Z$ using
$j=9=\{1,0,0,1\}$.
}
The diagonals of these $2^n$ basis operators correspond to the Walsh functions~\cite{WalshACS}, an orthogonal and complete set of basis functions with broad applicability in digital signal processing, particularly in image processing and compression.
In the binary-encoded ordering described in Eq.~\eqref{eq:opdef}, the associated Walsh functions correspond to rows in the natural-ordered WHT with unit matrix elements, ${\bf\rm H}^{(1)}_{n} = 2^{\frac{n}{2}} {\rm H}^{\otimes n}$ with ${\rm H} = \frac{1}{\sqrt{2}}\begin{pmatrix}
  1 & 1 \\
  1 & -1
\end{pmatrix}$
the standard single-qubit Hadamard operator.
In application,
this basis is related to transformations that are designed to calculate controlled rotation angles for the quantum circuit preparation of an arbitrary real wavefunction through sequential Hilbert space bisection~\cite{Klco:2019xro,Klco:2019yrb,Klco:2020aud} (now available in IBM's Qiskit~\cite{normaldistributionibm,quantumcomputingUKtutorial}).
The WHT  organizes a basis by  \emph{sequency} contributions,
or by the number of digital zero crossings,
displaying hierarchies that can be leveraged
for computational resource reduction,
as schematically shown in Fig.~\ref{fig:diagram}.
In Fig.~\ref{fig:diagram}, sequency modes are presented next to the most-UV logical qubit responsible for their excitation.

If the relevant wavefunction for expectation values or the operator itself is localized,
it may be advantageous to work in an alternate basis of Haar wavelets~\cite{Haar1910}.
Modifying the $\mathcal{O}_j$'s to capture local structure at decreasing length scales
(rather than the global structure of Walsh functions), the Haar basis reduces the number
of contributing basis operators that need to be considered.
However, this choice of operator basis will not affect the sensitivity of the expectation value to
noise in each logical qubit, which is the relevant feature for HI-QEC.

\section{Tracking Hierarchies Through Depolarization}
In order to track the propagation of intrinsic hierarchies through individual qubit contributions to the overall precision of a quantum simulation, we consider a simple single-qubit depolarizing noise channel.
In the presence of this channel, $\mathcal{E}$, a density matrix, $\rho$, is transformed as,
\begin{align}
  \mathcal{E}(\rho) &= \sum_{\mu=0}^3 K_\mu \rho K_\mu^\dagger
 \ \ \ ,
  \nonumber \\
  K_0 &= \sqrt{1-\eta}\ \mathbb{I},\ K_j = \sqrt{\frac{\eta}{3}} \sigma_j \ \ ,
  \label{eq:depolarizingkraus}
\end{align}
with $\eta$ the probability that the state is modified by isotropic Pauli noise.
Note that with a redefinition of $\eta \rightarrow \eta' =  \frac{3}{4} \eta$, $\eta'$ is the probability that the state is replaced by the maximally mixed state.
Expectation values of the operators defined in Eq.~\eqref{eq:opdef} under the effects of depolarizing noise on qubit $q$
have a conveniently constrained response,
\begin{align}
  \langle\mathcal{O}\rangle(\eta, q)
  &= \sum_{\mu=0}^3 \Tr\left[ K_\mu^{(q)} \rho \left(K_\mu^{(q)}\right)^\dagger \mathcal{O}\right]
   \ \ \ ,
   \nonumber \\
  & = \sum_\mu\sum_{j = 0}^{2^n-1} \beta_j \Tr\left[\rho \left(K^{(q)}_\mu\right)^\dagger \mathcal{O}_j K^{(q)}_\mu\right]
  \ \ \ ,
\end{align}
where the cyclic property of the trace has been used to emphasize that the noise channel can be considered to act upon the density matrix or the observable basis.
In this basis of observables, the Kraus operators, $K_\mu$, of the depolarizing channel simply produce
a negative sign in the $\mathcal{O}_j$ contribution if $\mu = \{1,2\}$ and $j_q = 1$
(i.e., for the $X,Y$ Kraus transformations of the $\mathcal{O}_j$'s with a $Z$ at the location of the depolarizing qubit),
\begin{multline}
  \langle \mathcal{O} \rangle (\eta, q) = \sum_{j = 0}^{2^n-1} \beta_j \langle \mathcal{O}_j \rangle
  \\
  \times\left[\left( 1 - \frac{2\eta}{3}\right) +  (-1)^{\delta_{j_q,1}}  \frac{2\eta}{3}\right]
  \ \ \ .
  \label{eq:singlequbitdepolarizing}
\end{multline}
If $\mathcal{O}_j $ contains an identity at the location of the depolarizing
qubit, $j_q=0$ and the $j^{\text{th}}$ contribution to the expectation value is insensitive to the noise impacting that qubit.
On the other hand,
if the basis operator contains a $Z$ at the location of the depolarized qubit, $j_q = 1$ and the $j^{\text{th}}$ contribution to the expectation value gains a multiplicative factor of $\left(1-\frac{4\eta}{3}\right)$.
Importantly, the magnitude of this multiplicative factor is not dependent on which qubit is affected by noise
beyond its binary sensitivity determined by the symmetry of the $\mathcal{O}_j$,
i.e., every noise-sensitive qubit will affect $\langle \mathcal{O}_j\rangle$ in the same way.

\section{Hierarchical Noise Sensitivity}
It follows from Eq.~\eqref{eq:singlequbitdepolarizing} that
when multiple qubits experience independent depolarizing noise,
the expectation value of diagonal operators can be written as,
\begin{multline}
 \langle \mathcal{O} \rangle (\boldsymbol{\eta}) =
 \sum_{j = 0}^{2^n-1} \beta_j
 \langle \mathcal{O}_j \rangle \\*
 \times
  \prod_q\
 \left[\left( 1 - \frac{2\eta_q}{3}\right) +  (-1)^{\delta_{j_q,1}}  \frac{2\eta_q}{3}\right]
 \ \ .
 \label{eq:expectnoiseA}
  \end{multline}
Any hierarchies present in the quantum system
that may be  leveraged to reduce the physical-to-logical error correction requirements reside in
 1.) $\beta_j$, the projection of the chosen observable in the $\mathcal{O}_j$ basis and
 2.) $\langle \mathcal{O}_j \rangle$,
 the wavefunction-dependent expectation values.

The first source of hierarchy, the distribution of $\vec{\beta}$, is determined only by the form of the operator.
Low-sequency probes of the system (corresponding to long-wavelength probes in the simulated system if logical qubits have been mapped hierarchically) tend to be weighted toward smaller reversed-bitstring $j$ values, while the converse is true for high-sequency probes.
The matrix element is sensitive through the basis operator to the depolarization present on individual qubits.
 For example,
 in the case of an operator that is independent of the orientation of any of the qubits,
 $\mathcal{O}_0$ ,  the expectation value is independent of depolarizing noise
 $\langle\mathcal{O}_0\rangle \rightarrow \langle\mathcal{O}_0\rangle $.
 In contrast, the operator that depends on the alignment of all  qubits, $\mathcal{O}_{2^n-1} $,
 suffers maximal suppression due to depolarization,
 $\langle\mathcal{O}_{2^n-1}\rangle \rightarrow  \left(\prod_{q} \left(1-\frac{4\eta_q}{ 3}\right)\right) \langle\mathcal{O}_{2^n-1} \rangle$.
 The noise-scaling of  other operators lies between these two extremes, with sensitivity generally increasing with the number of \enquote{1}s in the binary representation of~$j$.

For the second source of hierarchy, the wavefunction dependence of the matrix elements, it is the square decomposition of the wavefunction that
contributes the relevant intrinsic structure.
Writing
$|\psi \rangle= \sum\limits_{\ell=0}^{2^n -1} \psi_\ell\ |\ell\rangle$, where $|\ell\rangle$ is a computational basis state with unit amplitude only at position $\ell$,
the matrix elements of the
$\mathcal{O}_j$ relevant for  Eq.~(\ref{eq:expectnoiseA})
are
\begin{align}
\langle \mathcal{O}_j \rangle\ & = \
\langle\psi|\
\mathcal{O}_j
\ |\psi\rangle
 =
 \sum_\ell \ |\psi_\ell |^2\ (-1)^{ j \cdot \ell} \ \ \ , \nonumber \\
 &
 \ =\
 \left[\  {\bf\rm H}^{(1)} . \boldsymbol{\psi}^2 \right]_j
 \ \ ,
 \label{eq:Ohada}
  \end{align}
where $j \cdot \ell = j_0 \ell_0 + j_1 \ell_1+... + j_{n-1}\ell_{n-1} $ is the inner product of the binary representations
and $\boldsymbol{\psi}^2 = \left\{ |\psi_\ell |^2\right\}$ is the vector of probabilities.
The convenient property allowing the above succinct expression of the expectation values is that the rows of the natural-ordered WHT produce the diagonals of the $\mathcal{O}_j$'s as defined in Eq.~\eqref{eq:opdef}.

\begin{figure}
  \includegraphics[width = 0.9\columnwidth]{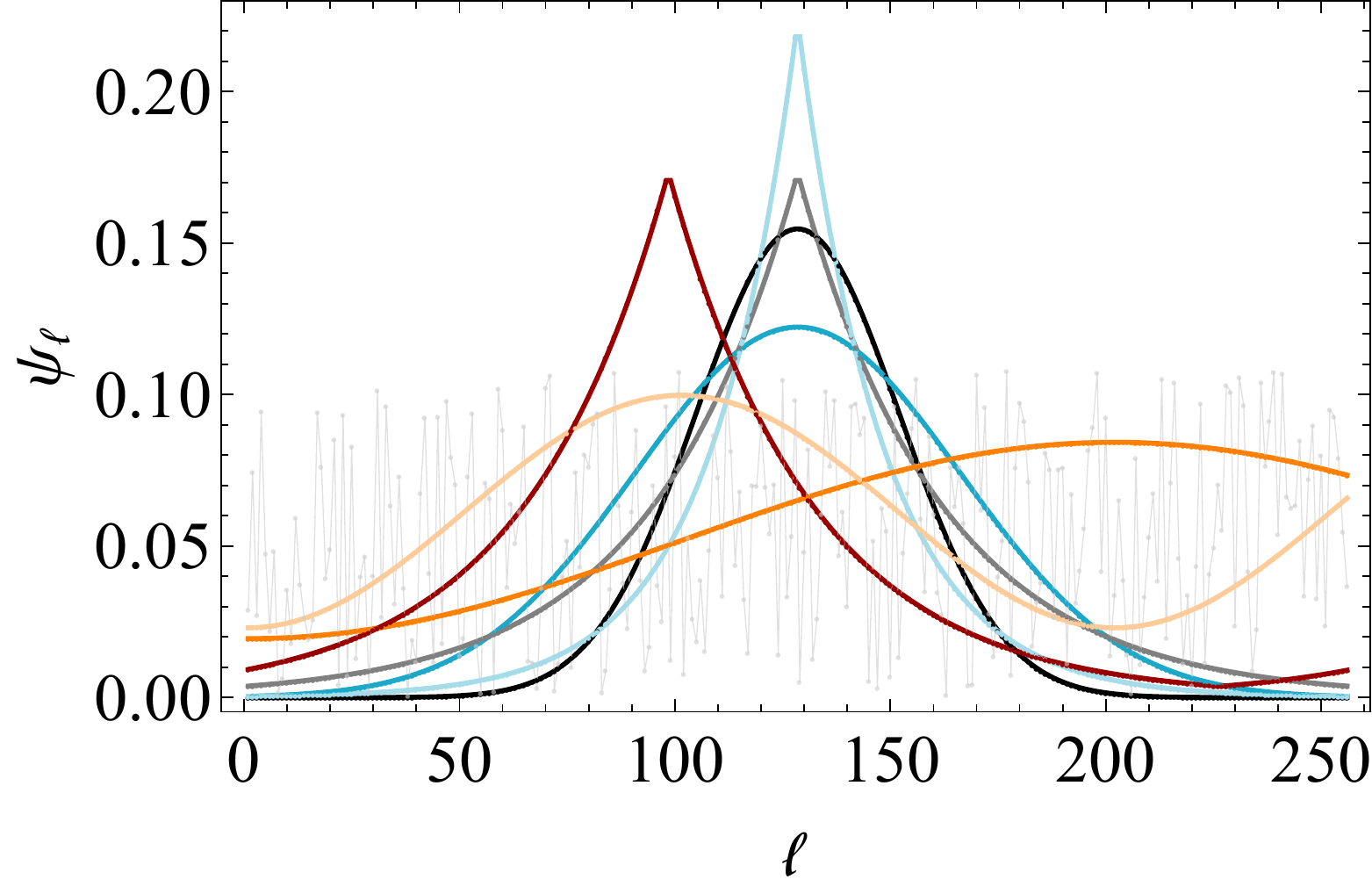}
  \includegraphics[width = 0.9\columnwidth]{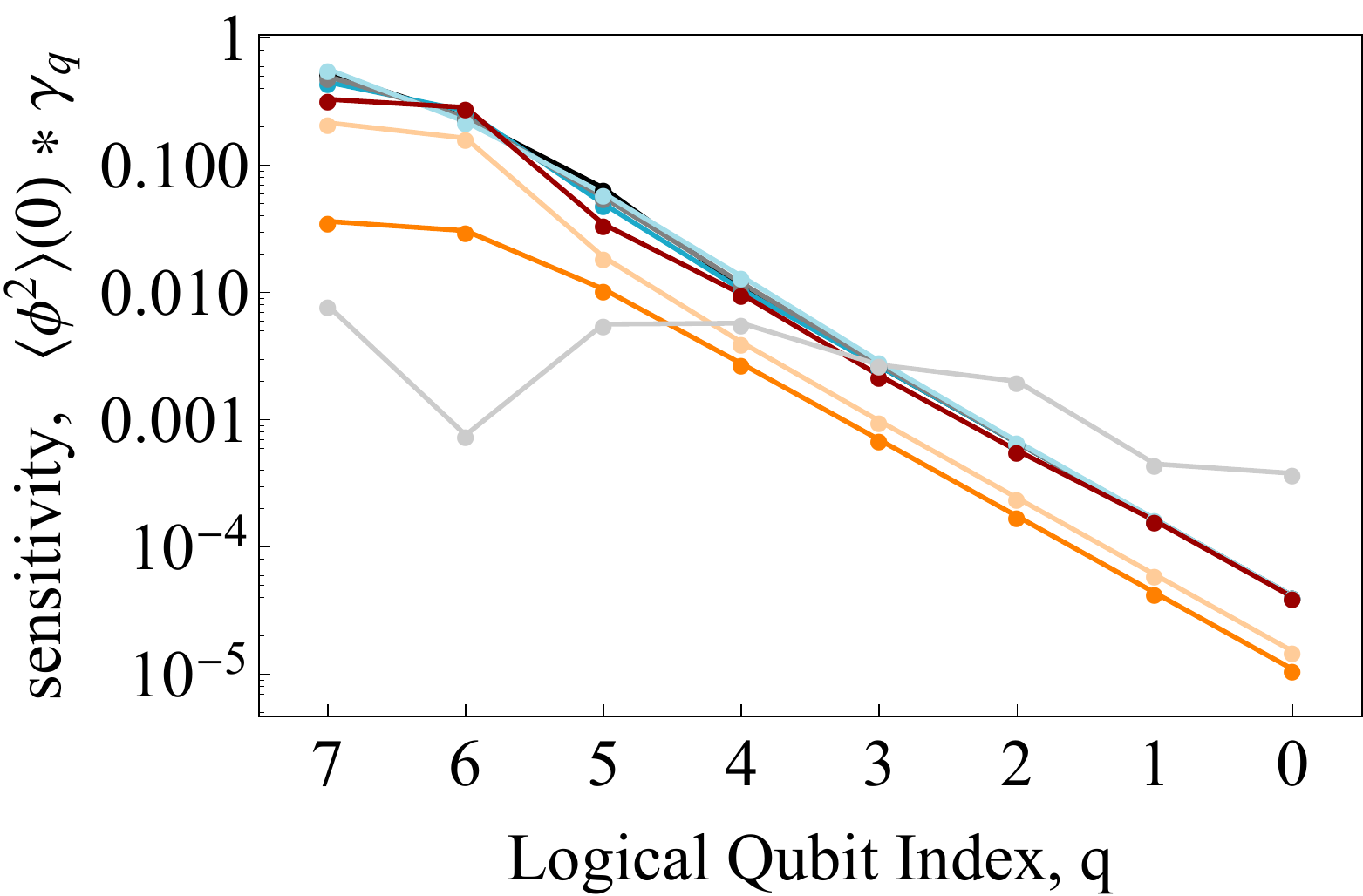}
  \caption{(lower) Noise sensitivities as a function of logical qubit index for (upper) a collection of smooth wavefunctions and one randomly generated wavefunction, digitized in the (hierarchical) binary basis.}
  \label{fig:gammas}
\end{figure}

Demonstrating these two hierarchical ingredients in a small system representing a building block for quantum simulation of lattice field theories, Appendix~\ref{app:fourqubits} gives an example of four qubits supporting a digitally sampled Gaussian wavefunction across the Hilbert space.
This example uses the binary (JLP) basis, which exhibits a hierarchy in energy scales across logical qubits.
The effects of depolarizing noise are determined for simple observables probing decreasing wavelengths, illustrating the heterogeneous distribution of logical qubit noise sensitivities.
For example, the logical qubit sensitivities in the matrix element of $\phi^2$ are
\begin{eqnarray}
\langle \phi^2 \rangle ({\bm\eta}) & = &
\langle \phi^2  \rangle (0)  \left[\
1\ +\ 0.094\eta_0 + 0.378 \eta_1
\right.
\nonumber\\
&& \left. \qquad
+ 2.151\eta_2  + 2.890 \eta_3
+ {\cal O}({\bm\eta}^2)
\ \right]
   \ .
   \label{eq:phi2gammas4}
\end{eqnarray}
The contribution to the error from depolarizing noise experienced by
qubit-3 (responsible for IR) is $\sim \times 30$ that from qubit-0 (UV).
Consequently, there is an inverted
hierarchy in the minimum logical-qubit fidelities necessary to provide a determination of long-wavelength observables to fixed precision.
Such a hierarchy can be leveraged through a heterogeneous distribution of physical qubits in the construction of logical qubits.
In general, observables considered in this work distorted by depolarizing noise take the form,
\begin{equation}
 \langle \mathcal{O} \rangle ({\bm\eta})  =
 \langle \mathcal{O} \rangle (0) \left[\ 1\ +\ \sum_{q=1}^n \gamma_q \eta_q \  +\ O({\bm\eta}^2) \right]
\ \ \ ,
\label{eq:opeta}
\end{equation}
where the $\gamma_q$ are observable and wavefunction dependent coefficients capturing the sensitivity of the expectation value to noise on each logical qubit to linear order in the logical error rates.
Extending the example of the logical qubit sensitivities for $\langle \phi^2\rangle$ in a smooth Gaussian wavefunction to eight qubits with Gaussian parameters $\mu = \frac{2^n-1}{2}, \sigma = 50/3$, one finds $\vec{\gamma} = \{31.15, 13.91, 3.86, 0.64, 0.15, 0.038, 0.0096, 0.0024 \}$.
As shown in Fig.~\ref{fig:gammas}, these logical qubit senstivities are exponentially suppressed for UV qubits, $\gamma_q \sim e^{\xi q}$, a general feature expected in the long-wavelength expectation values of smooth wavefunctions.

\section{HI-QEC}

Because proposals for QEC achieve remarkable scaling---exponential improvements in logical qubit fidelity with the addition of a polynomial number of physical qubits---application-specific improvements or specializations to the QEC scheme are usually expected to have little impact on quantum resource requirements.
However, the exponential suppression in the sensitivity of
low-energy observables to noise impacting UV logical qubits allows the opportunity for a non-trivial multiplicative reduction in physical qubit requirements.
When Hilbert spaces are hierarchically organized onto logical qubits, individual fidelities can be optimized through HI-QEC.

To gain a quantitative estimate of the impact of such hierarchies on QEC resources,
we consider the {\tt surface code}~\cite{Kitaev_2003,Gottesman:1997zz,freedman2001projective,Bravyi:1998sy,Dennis_2002,Wang:2002ph} as a concrete mapping from physical to logical qubits.
In the {\tt surface code}, quantum information for computation is stored non-locally in a lattice of physical qubits and protected from local sources of noise as a result of  logical operations being topologically non-trivial.
In each cycle of the code, local stabilizers are measured throughout the volume and classical algorithmic techniques are utilized to determine the most probable error diagnosis.
With increasing surface size and/or decreasing error rates, resulting correction procedures become less likely to inadvertently produce a topologically non-trivial excitation of the lattice, allowing detected quantum errors to be successfully corrected or classically compensated for through redefinition of measurement and operator bases.
For a physical qubit per-step error rate, $p$, below a threshold value $p_{\rm th}$, the fidelity of the logical qubit increases exponentially with increasing surface size~\cite{548464,Knill_1998,Kitaev_2003,aharonov2008fault}.
Detailed studies have been performed of the associated error thresholds in the  {\tt surface code}, including depolarization errors, see e.g., Ref.~\cite{Bombin_2012,PhysRevA.86.032324}.
The per-cycle logical qubit X-error rate has been empirically (through classical simulation) determined to be~\cite{PhysRevA.86.032324},
\begin{equation}
  P_L \sim c_0 \left(p/p_{\rm th}\right)^{\left\lfloor(d+1)/2 \right\rfloor} \qquad p < p_{\rm th} \ \ \ ,
\end{equation}
where $d$ is the side-length of the physical
qubit array forming the {\tt surface code}, $c_0\sim 0.03$, and $p_{\rm th}=0.0057$ (for additional details see Fig. 4 in Ref.~\cite{PhysRevA.86.032324}).
For the purposes of this work, we will assume $p$, and thus $P_L$, is time independent,
both from external backgrounds and the application of quantum circuits for computation.

Assuming a logical-qubit depolarizing error rate that is isotropic among the Pauli bases, the per-cycle logical error rate for the $q^{\rm th}$ logical qubit, $P_L^{(q)}$,
is related to the depolarizing noise parameter of Eq.~\eqref{eq:depolarizingkraus} as $\eta_q = P_L^{(q)} N_{\rm cycles}$.
In this framework, the number of error correction cycles, $N_{\rm cycles}$, is related to, among a number of factors, the desired depth of circuits providing reliable calculations.
Requiring that the matrix element of the operator is fractionally within $\epsilon$ of its true value
and neglecting higher order contributions in Eq.~(\ref{eq:opeta}),
yields
\begin{eqnarray}
\sum_q\ \gamma_q P_L^{(q)} & \le & \frac{\epsilon}{ N_{\rm cycles}}
\ \ \ .
\end{eqnarray}
Assuming uniform physical error rates below $p_{\rm th}$, the construction of {\tt surface codes} with heterogeneous physical-to-logical ratios can be informed by the constraint that all error contributions to a chosen observable are uniform across the logical qubits,
\begin{align}
  \gamma_q P_L^{(q)} &\le  \frac{\epsilon}{ n N_{\rm cycles}} \qquad \forall \ q \in \{0,\ldots, n-1\} \ \ \ ,
  \nonumber\\
   d_q \ & \gtrsim 2\left\lceil \frac{  \log\left( \frac{\epsilon}{ n\overline{c}_0 \gamma_q} \right)  }{  \log\left(  \frac{p}{ p_{\rm th}}\right)} \right\rceil-1
\ \ \  ,
\label{eq:uniformErrLogicalConfiguration}
\end{align}
where $ \overline{c}_0 = N_{\rm cycles} c_0 $,
and the number of qubits defining the $q^{\rm th}$ logical qubit is $d_q^2$.
The logarithmic relation between $d_q$ and $\gamma_q$ means that optimizing in the number of physical qubits for each logical qubit results in a polynomial reduction in the number of physical qubits.
In particular, the exponential decay of $\gamma_q$ leads to a linear dependence of $d_q$ on the logical qubit index, a surface effect on each logical {\tt surface code}.

\begin{figure}
  \includegraphics[width = 0.9\columnwidth]{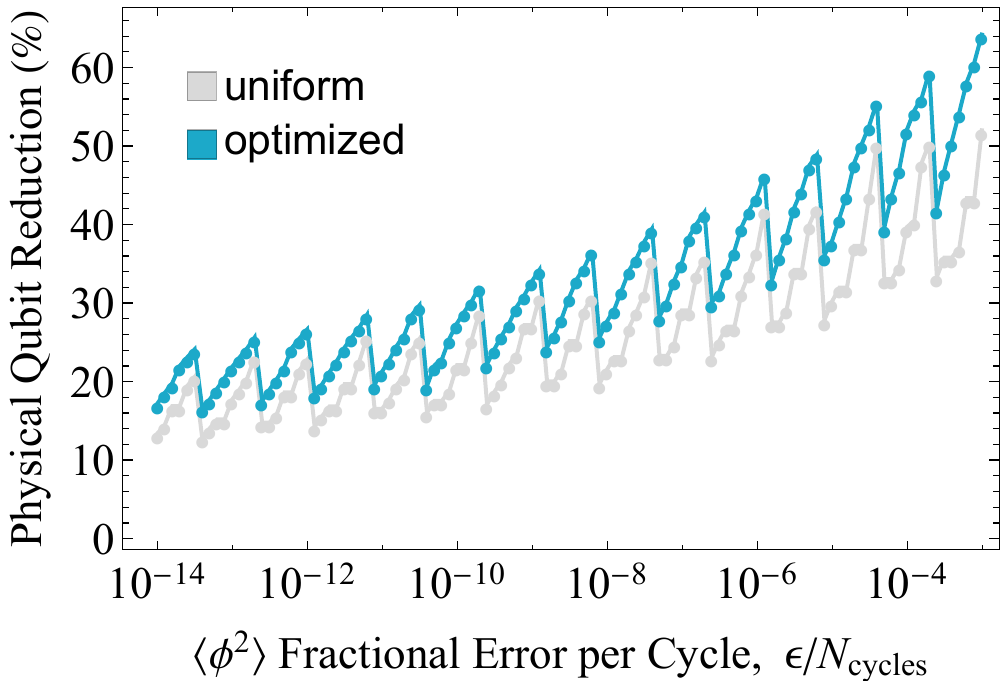}
  \caption{The percent reduction, through HI-QEC, in physical qubits necessary to achieve per-cycle fractional precision, $\epsilon/N_{\rm cycles}$, in the expectation value of $\phi^2$ (i.e., after $N_{\rm cycles}$ of the {\tt surface code}, the fractional error in $\langle \phi^2\rangle$ is $\epsilon$).
  Results are shown for heterogeneous physical-to-logical qubit distributions that are (gray) constrained to uniformly distribute error among logical qubits as described in Eq.~\ref{eq:uniformErrLogicalConfiguration} and (blue) optimized without constraint.
  }
  \label{fig:physqubitreduction}
\end{figure}
Consider the eight-qubit system discussed at the end of the previous section of a Gaussian wavefunction with $\sigma = 50/3$, with a physical qubit error rate of
$p=10^{-3}$ for a target fractional error per cycle of the {\tt surface code} of
$\epsilon/N_{\rm cycles}=10^{-5}$.
If all logical qubits are constructed to have the same error rate,
\begin{equation}
  d \gtrsim 2 \left\lceil \frac{\log\left( \frac{\epsilon}{\overline{c}_0 \sum_q \gamma_q} \right) }{\log \left( \frac{p}{p_{\rm th}}\right) }  \right\rceil -1 \ \ \ ,
\end{equation}
the required surface size for each logical qubit is $d = 13$, demanding a total of 1352 physical qubits.
Distributing error contributions uniformly among the logical qubits, as discussed in Eq.~\eqref{eq:uniformErrLogicalConfiguration}, leads to a heterogeneous distribution of physical qubits among the logical qubits as $\mathbf{d} = \{15, 15, 13, 11, 9, 7, 7, 5\}$ for a total of 944 physical qubits achieving a fractional error rate per cycle of $4.9*10^{-6}$.
Further optimization can be performed that removes the restriction on uniformly distributed error contributions, resulting in an optimal physical qubit distribution of $\mathbf{d} = \{15, 13, 11, 11, 9, 7, 7, 5\}$ for a total of 840 qubits achieving a fractional error rate per cycle of $9.4*10^{-6}$.
This final optimization allows error contributions to reflect the projective importance of each qubit in the observable of interest, in this case the low-energy observable $\langle\phi^2 \rangle$.
By organizing the scale information into qubit-wise hierarchies, the number of physical qubits required in this simple example can be reduced by $ \sim 37\%$.
Figure~\ref{fig:physqubitreduction} presents the percent reduction in physical qubits for this eight-qubit system throughout a range of desired per-cycle fractional error rates.~\footnote{The integer nature of {\tt surface code} size produces a natural discretization in achievable fidelities.
The sawtooth behavior in Fig.~\ref{fig:physqubitreduction} is dominantly due to the physical qubit requirements of the homogeneous distribution exhibiting large changes relative to those of heterogeneous distributions.
With higher fidelity resolution afforded by HI-QEC, the relationship between physical qubit number and fidelity is \enquote{smoothed}.}
For a per-cycle fractional error of $\langle \phi^2\rangle$ equal to the error rate of the physical qubits $(p = 10^{-3})$, up to 60\% reductions in physical qubits are found.
As the target error decreases, the fidelities of UV qubits become of greater impact and improvements through heterogeneous distribution are diminished, though remain $\sim 20\%$ even near per-cycle fractional error rates around classical machine precision.
These results indicate tens of percent reductions in number of physical qubits, and specifically not in the dimension of the Hilbert space.
Heterogeneous distributions of physical qubits creating an optimal system of logical qubits informed by hierarchies intrinsic to the
physical system and qubit mapping
are expected to profitably impact near-term simulations, providing non-zero but diminishing returns as precisions improve.

\section{Conclusions}

Toward the quantum simulation of Standard Model physics,
we consider the importance of representing physical Hilbert spaces
on quantum degrees of freedom in a way that manifests
underlying hierarchies.
This manifestation allows bands of hierarchical quantities to be dominantly supported by particular qubits.
When considering momenta and length scales as the hierarchical quantities of interest,
a hierarchical mapping allows
qubits to be located on a UV-IR spectrum,
where UV and IR qubits dominantly capture the structure of high and low energies, respectively.
As such,  observables with support in localized energy bands may be calculated
more efficiently by leveraging a similarly localized distribution of logical qubit fidelities,
potentially
extending the scientific scope of an environment with constrained quantum resources.

In this work,
and with an eye toward the simulation of QFTs and their low-energy EFTs,
we demonstrate the practical utility of hierarchical qubit mappings with a focus on
smoothly varying wavefunctions across a single Hilbert space and low-energy observables.
Explicitly, the binary (JLP) basis used to map scalar QFTs onto quantum registers
provides an organizational principle among qubits aligning with sequency.
We show that this inherent relative importance of qubits in the sequency hierarchy can be advantageously
utilized in the computation of matrix elements of low-energy observables in smoothly varying wavefunctions.
The single-qubit isotropic depolarizing noise model is used to track the sensitivity of diagonal matrix elements to noise contributions from each qubit.
Modulo the presence of strong out-of-band, multi-qubit correlated noise contributions, e.g., between UV and IR qubits, this is expected to be a representative noise model for probing the propagation of hierarchies.
The linear response to depolarizing noise is found to scale exponentially with qubit number for a wide range of smooth wavefunctions.
As such, the sequency hierarchy is illustrated to give rise to strongly non-uniform noise sensitivities,
and thus motivates structured heterogeneity in  distributions of logical qubit fidelities.
Using the simulated properties of the {\tt surface code} as one explicit map between physical and logical qubits,
we quantify the impact of  this heterogeneity on the number of physical qubits required to determine low-energy observables to fixed precision.
For  scientifically impactful quantum calculations of nuclear and particle physics,
where early precision goals will be at the few to tens of percent level,
the reduction in physical qubits that  can be achieved through co-design
is anticipated to be significant when hierarchies are  identified, mapped and implemented.

While this work has focused on the physical-to-logical Hilbert space requirements for QEC in qubit systems, hierarchical maps can be advantageously utilized in the NISQ era (without error correction) when physical qubits exhibit non-uniform fidelities.
Beyond qubits, mapping hierarchies in Hilbert spaces generically provides the opportunity for optimizing fidelities among bands of states and the operations within.
In qudits, e.g., SRF cavity systems under development at LLNL and FNAL~\cite{Romanenko:2018nut,Holland:2019zju}, the presence of hierarchies reduces the number of states required to be controlled with high precision.

In quantum simulations, any practical mapping involves the action of both diagonal and non-diagonal operators, e.g., the electric (Casimir) and magnetic (plaquette) operators in the Kogut-Susskind Hamiltonian describing SU(N) lattice gauge theories~\cite{PhysRevD.11.395}.
The diagonal operator hierarchies discussed in this work are expected to persist for band-diagonal operators, e.g., the
SU(3) plaquette operator with two-dimensional locality in the  hexagonally connected color space, and can be utilized in the co-design of QCD quantum simulations.
However, quantifying the propagations of such hierarchies and the subsequent quantum resource optimizations for lattice gauge theory simulations remains to be performed.

With dynamical, real-time simulations, one can easily imagine the future utility of spatially local physical-to-logical qubit re-distributions optimized through an iterative workflow of tuning and production.
After low-fidelity tuning simulations with a uniform distribution of physical-to-logical qubits identifying relevant Hilbert space support, static hierarchical redistributions of qubits will improve the fidelity of subsequent simulations.
The computation of S-matrix elements in QFTs from initially localized wavepackets or jet production in heavy-ion collisions  provide concrete examples in which precision can be enhanced along trajectories of particle fluxes.
This local redistribution allows quantum resources to preferentially scale with the integrated volume of wavefunction support rather than the lattice volume.
More speculatively, dynamically evolving the allocation of physical qubits to track the propagating wavepackets in real-time throughout the scattering process would allow quantum resources to preferentially scale with the size of the wavepacket as opposed to the scattering volume.
This is analogous to the use of adaptive multi-grid in classical simulations, though requires further consideration of engaging with repartitions of non-local quantum information.

\begin{acknowledgments}
NK is supported in part by the Walter Burke Institute for Theoretical Physics,
and by the U.S. Department of Energy Office of Science,
Office of Advanced Scientific Computing Research, (DE-SC0020290), and Office of High Energy Physics DE-ACO2-07CH11359.
MJS is supported in part by the U.S. Department of Energy, Office of Science, Office of Nuclear Physics,
InQubator for Quantum Simulation (IQuS) under Award Number DOE (NP) Award DE-SC0020970 through the
Quantum Horizons: QIS Research and Innovation for Nuclear Science Initiative.
Results of numerical calculations presented within this manuscript are available upon email request.
\end{acknowledgments}

\appendix
\renewcommand{\thefigure}{A\arabic{figure}}
\renewcommand{\theHfigure}{A\arabic{figure}}
\setcounter{figure}{0}
\renewcommand{\thetable}{A\arabic{table}}
\renewcommand{\theHtable}{A\arabic{table}}
\setcounter{table}{0}

\section{Low-Energy Example}
\label{app:fourqubits}

\begin{figure}[!ht]
\centering
\includegraphics[width = 0.4\textwidth]{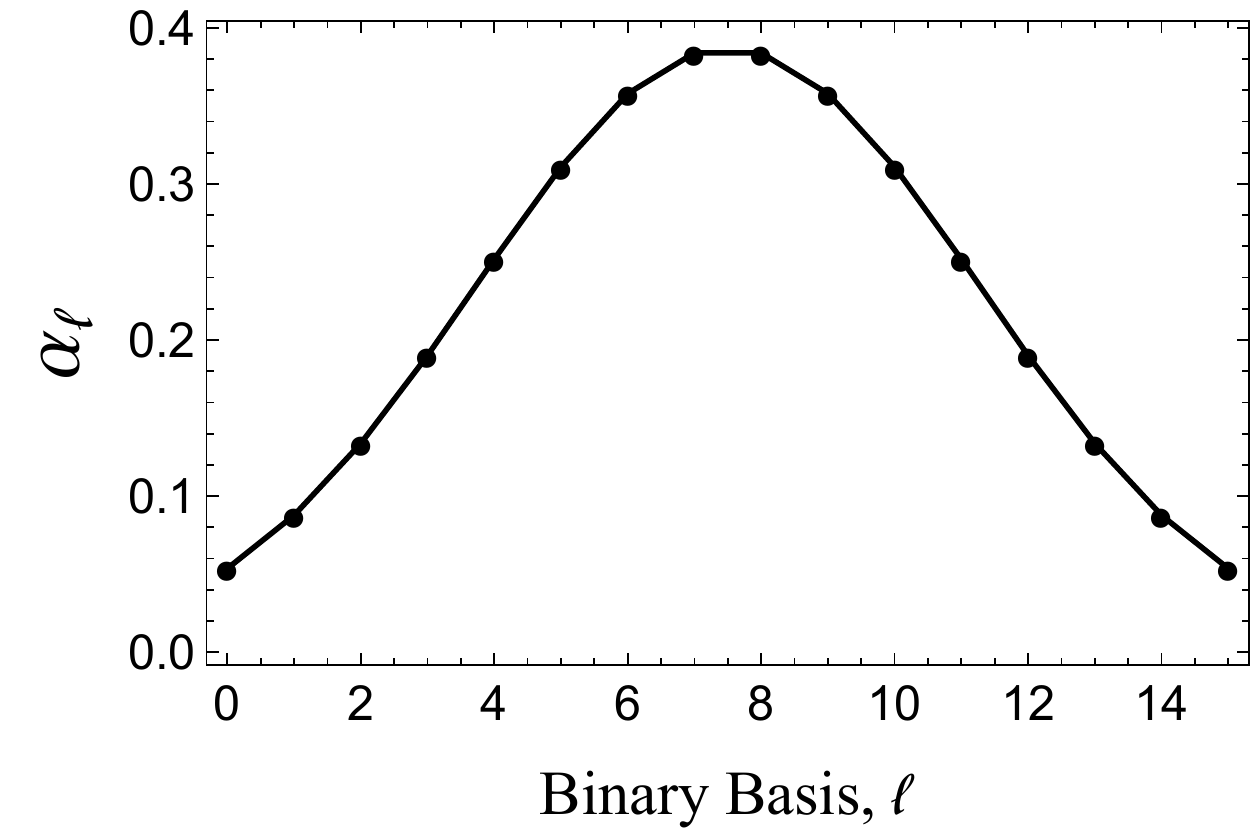}
\includegraphics[width = 0.4\textwidth]{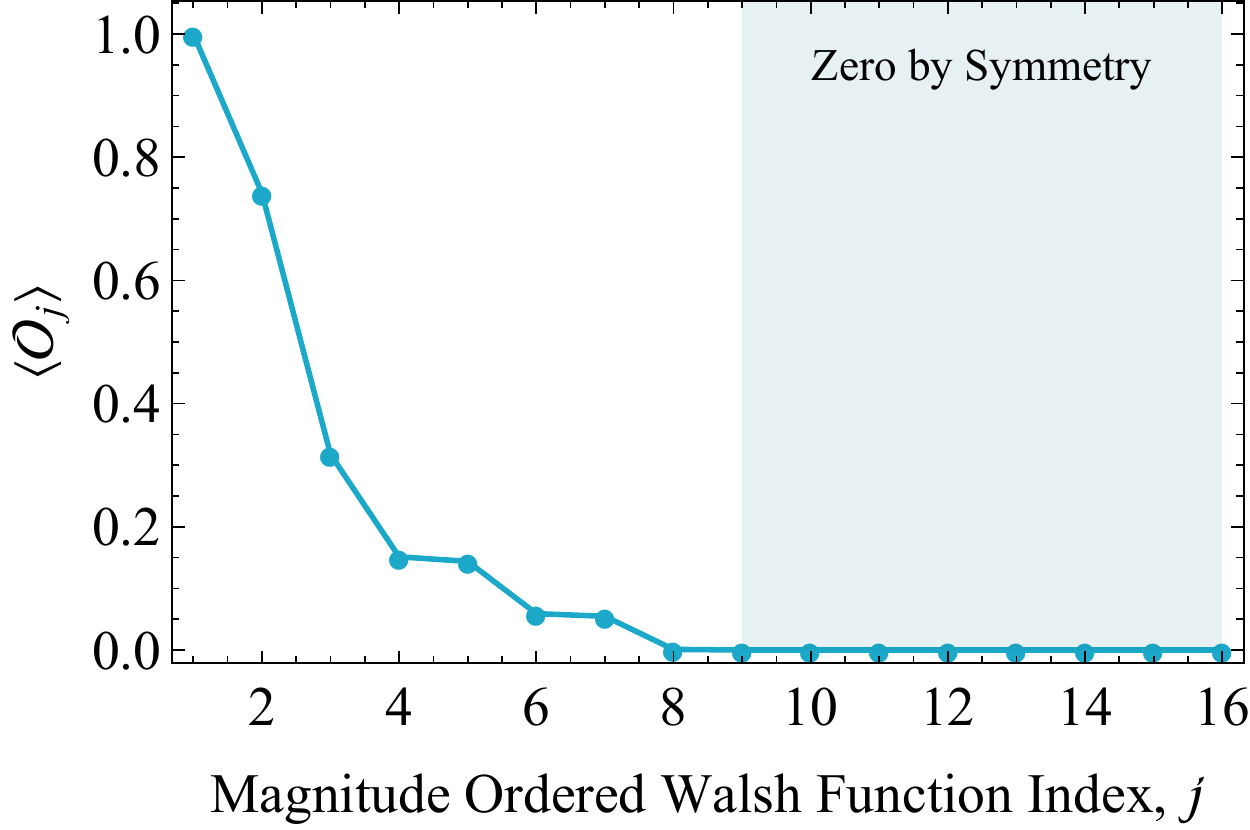}
  \caption{
  (upper)
  Probabilities, $|\psi_\ell|^2$, associated with the Gaussian wavefunction
  prepared on four qubits with the parameters defined in
  Eq.~(\ref{eq:GaussProbs}).
  (lower)
  The expectation values of $\mathcal{O}_j$, calculated using the natural-ordered WHT of Eq.~(\ref{eq:Hada}) and sorted by magnitude.}
  \label{fig:GaussProbs}
\end{figure}

This appendix illustrates through example the logical qubit noise propagation in the calculation of low-energy expectation values.
We focus on a single lattice site of non-interacting scalar field theory ground state,
i.e., a Gaussian,
digitized over a 16-dimensional Hilbert space of four qubits ($n = 4$).
The normalized Gaussian wavefunction is
\begin{align}
\psi_\ell   &= \mathcal{N} \exp \left[ -\frac{1}{4} \left( \frac{\ell-\mu}{\sigma} \right)^2 \right] \ \ \ , \nonumber \\
\mathcal{N} &= \frac{1}{\sqrt{\sum_\ell |\psi_\ell|^2}}\ \ \   ,
\label{eq:GaussProbs}
\end{align}
where $\sigma=8/3$  and $\mu = 15/2$ are chosen to provide the amplitudes shown in Fig.~\ref{fig:GaussProbs}.
The four-qubit, natural-ordered WHT in Eq.~(\ref{eq:Ohada}) is given by
\begin{equation}
{\rm H}_4^{(1)}  =  \resizebox{.7\hsize}{!}{$
\left(
\begin{array}{cccccccccccccccc}
 1 & 1 & 1 & 1 & 1 & 1 & 1 & 1 & 1 & 1 & 1 & 1 & 1 & 1 & 1 & 1 \\
 1 & -1 & 1 & -1 & 1 & -1 & 1 & -1 & 1 & -1 & 1 & -1 & 1 & -1 & 1 & -1 \\
 1 & 1 & -1 & -1 & 1 & 1 & -1 & -1 & 1 & 1 & -1 & -1 & 1 & 1 & -1 & -1 \\
 1 & -1 & -1 & 1 & 1 & -1 & -1 & 1 & 1 & -1 & -1 & 1 & 1 & -1 & -1 & 1 \\
 1 & 1 & 1 & 1 & -1 & -1 & -1 & -1 & 1 & 1 & 1 & 1 & -1 & -1 & -1 & -1 \\
 1 & -1 & 1 & -1 & -1 & 1 & -1 & 1 & 1 & -1 & 1 & -1 & -1 & 1 & -1 & 1 \\
 1 & 1 & -1 & -1 & -1 & -1 & 1 & 1 & 1 & 1 & -1 & -1 & -1 & -1 & 1 & 1 \\
 1 & -1 & -1 & 1 & -1 & 1 & 1 & -1 & 1 & -1 & -1 & 1 & -1 & 1 & 1 & -1 \\
 1 & 1 & 1 & 1 & 1 & 1 & 1 & 1 & -1 & -1 & -1 & -1 & -1 & -1 & -1 & -1 \\
 1 & -1 & 1 & -1 & 1 & -1 & 1 & -1 & -1 & 1 & -1 & 1 & -1 & 1 & -1 & 1 \\
 1 & 1 & -1 & -1 & 1 & 1 & -1 & -1 & -1 & -1 & 1 & 1 & -1 & -1 & 1 & 1 \\
 1 & -1 & -1 & 1 & 1 & -1 & -1 & 1 & -1 & 1 & 1 & -1 & -1 & 1 & 1 & -1 \\
 1 & 1 & 1 & 1 & -1 & -1 & -1 & -1 & -1 & -1 & -1 & -1 & 1 & 1 & 1 & 1 \\
 1 & -1 & 1 & -1 & -1 & 1 & -1 & 1 & -1 & 1 & -1 & 1 & 1 & -1 & 1 & -1 \\
 1 & 1 & -1 & -1 & -1 & -1 & 1 & 1 & -1 & -1 & 1 & 1 & 1 & 1 & -1 & -1 \\
 1 & -1 & -1 & 1 & -1 & 1 & 1 & -1 & -1 & 1 & 1 & -1 & 1 & -1 & -1 & 1 \\
\end{array}
\right)
$}
\ \ \ .
  \label{eq:Hada}
\end{equation}
The operator basis defined in Eq.~(\ref{eq:opdef}) in binary order is,
\begin{align}
\left\{ \ \mathcal{O}_j \ \right\}
 =&
\{
IIII , IIIZ , IIZI , IIZZ ,
\nonumber \\
 &
\hspace{0.3cm}IZII , IZIZ , IZZI , IZZZ
\nonumber \\
 &
\hspace{0.3cm}ZIII , ZIIZ , ZIZI , ZIZZ ,
\nonumber \\
 &
\hspace{0.3cm} ZZII , ZZIZ , ZZZI , ZZZZ \}
\ \ \ .
\end{align}
Note the relation with the rows to the natural-ordered Hadamard of Eq.~\eqref{eq:Hada} as discussed in the main text.
Introducing noise parameters
${\bm\eta} = \{ \eta_3 , \eta_2 , \eta_1 , \eta_0 \}$,
the
expectation values of the operators in the presence of depolarizing noise, as shown in Eq.~(\ref{eq:expectnoiseA}),
in this state are
\begin{multline}
\langle \mathcal{O}_j \rangle
  \prod_q
 \left[\left( 1 - \frac{2\eta_q}{3}\right) +  (-1)^{\delta_{j_q,1}}  \frac{2\eta_q}{3}\right]
  = \\
  \left\{
  1.000
, \right. \\
0.000
, \\
0.000
, \\
-0.001+0.001 \eta_0+0.001 \eta_1-0.002 \eta_0 \eta_1
, \\
0.000
, \\
0.059-0.079 \eta_0-0.079 \eta_2+0.105 \eta_0 \eta_2
, \\
0.144-0.192 \eta_1-0.192 \eta_2+0.256 \eta_1 \eta_2
, \\
0.000
, \\
0.000
, \\
-0.151+0.202 \eta_0+0.202 \eta_3-0.269 \eta_0 \eta_3
, \\
-0.318+0.423 \eta_1+0.423 \eta_3-0.565 \eta_1 \eta_3
, \\
0.000
, \\
-0.742+0.989 \eta_2+0.989 \eta_3-1.320 \eta_2 \eta_3
, \\
0.000
, \\
0.000
, \\
0.055-0.073 \eta_0-0.073 \eta_1+0.097 \eta_0 \eta_1-0.073 \eta_2 \\ +0.097 \eta_0 \eta_2  +0.097 \eta_1 \eta_2-0.130 \eta_0 \eta_1 \eta_2 -0.073 \eta_3 \\ +0.097 \eta_0 \eta_3+0.097 \eta_1 \eta_3-0.130 \eta_0 \eta_1 \eta_3+0.097 \eta_2 \eta_3 \\ -0.130 \eta_0 \eta_2 \eta_3-0.130 \eta_1 \eta_2 \eta_3+ \left. 0.173 \eta_0 \eta_1 \eta_2 \eta_3
 \right\} \ .
   \label{eq:HadaNoise}
\end{multline}
Considering that the sequency associated with each binary-ordered $\mathcal{O}_j$ is $\{0, 15, 7, 8, 3, 12, 4, 11, 1, 14, 6, 9, 2, 13, 5, 10\}$, the second source of hierarchy discussed in the main text can be seen in the magnitudes of matrix element coefficients.
As discussed, the selection of active qubits generating uniform noise sensitivity in $\mathcal{O}_j$ is controlled by the distribution of 1's in the binary representation of~$j$.

\begin{table}
\footnotesize
\begin{tabular}{c|cc|cccccccc}
\hline
$j$ & $s$ & $q_s$ & $ \phi^2 $& $ \phi^4 $& $ \phi^6 $& $ \phi^8 $& $ \phi^{10} $& $ \phi^{12} $ & $ \phi^{14} $& $ \phi^{16} $ \\
\hline
0&0&-&0.378&0.256&0.205&0.178&0.161&0.151&0.144&0.139 \\
3&8&0&0.018&0.039&0.060&0.077&0.090&0.100&0.107&0.111 \\
5&12&0&0.036&0.070&0.085&0.093&0.100&0.105&0.110&0.113 \\
6&4&1&0.071&0.136&0.151&0.152&0.148&0.144&0.140&0.137 \\
9&14&0&0.071&0.079&0.087&0.094&0.100&0.105&0.110&0.113 \\
10&6&1&0.142&0.150&0.154&0.153&0.149&0.144&0.140&0.137 \\
12&2&2&0.284&0.240&0.202&0.177&0.161&0.151&0.144&0.139 \\
15&10&0&0.000&0.030&0.057&0.077&0.090&0.100&0.107&0.111 \\
\end{tabular}
   \caption{Non-zero values of $\beta_j$ for operators $\phi^p$ decomposed onto the four-qubit $\mathcal{O}_j$ basis operators. The column $s$ provides the sequency of $\text{diag}\left( \mathcal{O}_j\right)$, while the column $q_s$ indicates the most-UV qubit responsible for generating each sequency.  As explored in the text, for a qubit map hierarchy of energy scales, qubits capture the UV with decreasing index.}
  \label{tab:betavals}
\end{table}
\begin{figure}
  \includegraphics[width=0.9\columnwidth]{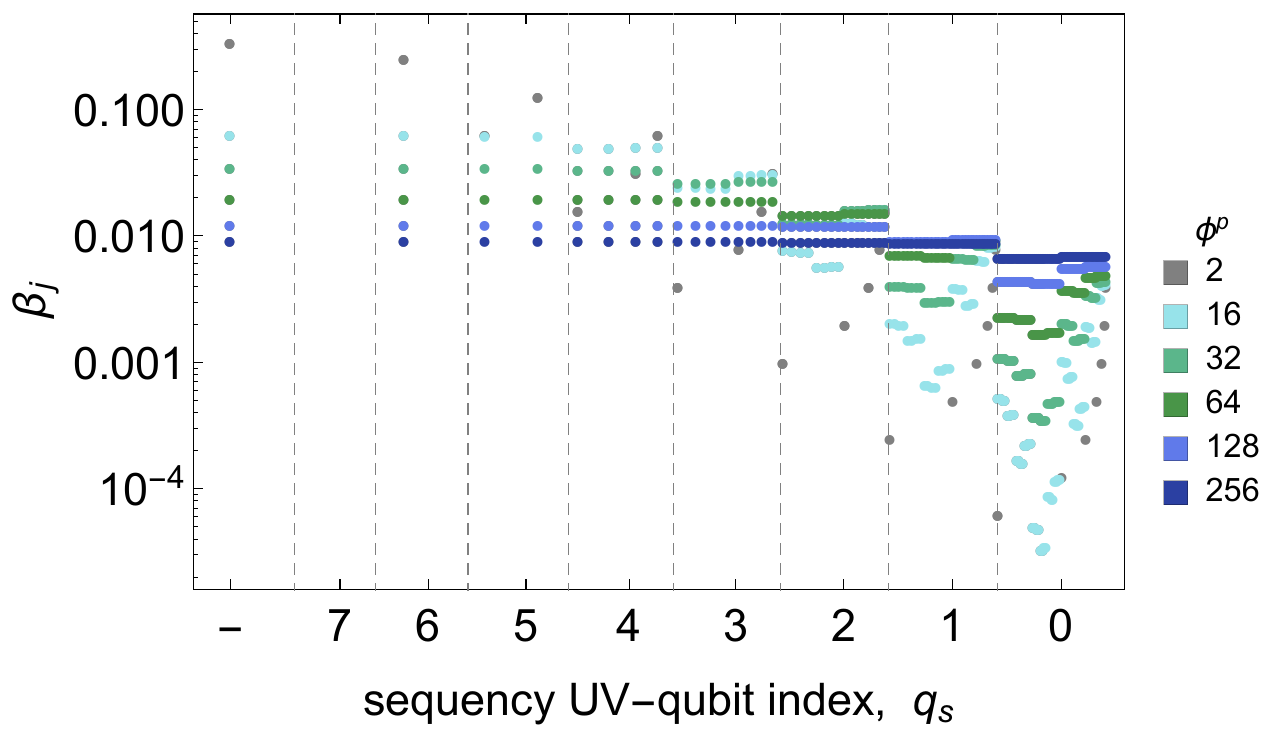}
  \caption{Basis operator projections, $\beta_j$, for powers of the operator, $\phi^p$, on eight qubits.
  The $\beta_j$'s are ordered by the sequency of ${\rm diag} \left(\mathcal{O}_j\right)$ and labeled by
  $q_s$, the most-UV qubit responsible for generating each sequency band.
  As explored in the text, for a qubit map hierarchy of energy scales, qubits capture the UV with decreasing index. }
  \label{fig:betasn8}
\end{figure}

To illustrate the first source of hierarchy discussed in the main text, we consider low moments of the operator, $\phi^p$.  In the scalar field application of this small example, this operator is the field operator.  When symmetrized about the origin in the binary (JLP) basis, the operator can be decomposed into single-qubit operators $\phi = \frac{1}{2^n -1} \sum\limits_{j = 0}^{n-1} 2^{j} \sigma^z_j$~\cite{Klco:2018zqz} for field values distributed between $\pm 1$ as
\begin{eqnarray}
\phi & = & \frac{1}{15}\left( IIIZ + 2 IIZI + 4 IZII + 8 ZIII \right)
  \nonumber\\
& = &
\frac{1}{15} \left( \mathcal{O}^{(\beta)}_1 + 2 \mathcal{O}^{(\beta)}_2 + 4 \mathcal{O}^{(\beta)}_4 + 8 \mathcal{O}^{(\beta)}_8 \right)
   \ \ \ .
\end{eqnarray}
When acting on the computational basis states, this field operator takes the values $\phi = \{\ 1, \frac{13}{ 15}, \frac{11}{ 15}, ......., \frac{1}{ 15}, -\frac{1}{ 15}, .... , -1\ \}$.   The operator decompositions of powers of $\phi$ are given in Table~\ref{tab:betavals} (only the non-zero projections are shown).
A hierarchy is apparent for low powers of $\phi$, organized again by sequency and by $q_s$, the most-UV qubit responsible for generating fluctuations at the scale of the sequency.
As the power of $\phi$ is increased, the hierarchy is systematically lost beginning in the IR.
This structure is more visible in Fig.~\ref{fig:betasn8}, showing results from an eight-qubit example that is
analogous to the four-qubit results in Table~\ref{tab:betavals}.
In these results, it is seen that the exponential decay in $\beta_j$ with sequency index for low powers of $\phi^p$ is systematically flattened in the IR as $p$ is increased.
Further, the hierarchy structure coincides with that found in the Hadamard transform of the Gaussian probability distribution for the selected parameters.
Both of these hierarchies, while distinct, result directly from working in a basis that naturally separates scales by digitized frequency.
The inner product of Eq.~(\ref{eq:HadaNoise}) with the vector of $\beta_j$'s in Table~(\ref{tab:betavals}) furnishes
\begin{eqnarray}
\langle \phi^2 \rangle ({\bm\eta}) & = &
\langle \phi^2  \rangle (0)  \left[\
1\ +\ 0.094\eta_0 + 0.378 \eta_1
\right.
\nonumber\\
&& \left. \qquad
+ 2.151\eta_2  + 2.890 \eta_3
+ {\cal O}({\bm\eta}^2)
\ \right]
   \ ,
\end{eqnarray}
(Eq.~\eqref{eq:phi2gammas4} of the main text) which shows the relative contributions,
from single-qubit depolarizing noise,
to the error in $\langle \phi^2 \rangle$.
Depolarization of qubits representing higher-digitization samples of the field (qubits 0,1) is less detrimental for this long wavelength operator in a smooth wavefunction.
As such, reducing the
error rates of qubits 2,3 is more impactful than reducing those of qubits 0,1.


\bibliography{biblio}

\end{document}